\documentclass[sn-mathphys,Numbered]{sn-jnl}

\usepackage{graphicx}%
\usepackage{multirow}%
\usepackage{amsmath,amssymb,amsfonts}%
\usepackage{amsthm}%
\usepackage{mathrsfs}%
\usepackage[title]{appendix}%
\usepackage{xcolor}%
\usepackage{textcomp}%
\usepackage{manyfoot}%
\usepackage{booktabs}%
\usepackage{algorithm}%
\usepackage{algorithmicx}%
\usepackage{algpseudocode}%
\usepackage{listings}%
\usepackage[utf8]{inputenc}
\DeclareUnicodeCharacter{2500}{ss}

\theoremstyle{thmstyleone}%
\theoremstyle{thmstyletwo}%

\theoremstyle{thmstylethree}%

\begin{document}

\title[Electromagnetic Multipoles for Morris-Thorne Wormhole]{Electromagnetic Multipoles for Morris-Thorne Wormhole}


\author[1,2]{\fnm{A. H.} \sur{Hasmani}}\email{$^2$ah\_hasmani@spuvvn.edu}

\author*[1,3]{\fnm{Sagar V.} \sur{Soni}}
\email{$^3$sagar.soni7878@gmail.com}

\author[1,4]{\fnm{Ravi} \sur{Panchal}}\email{$^4$ravipanchal1712@spuvvn.edu}

\affil[1]{\orgdiv{Department of Mathematics}, \orgname{Sardar Patel University}, \orgaddress{\city{Vallabh Vidyanagar}, \postcode{388120}, \state{Gujarat}, \country{India}}}


\abstract{Wormholes are interesting space-time structures connecting two asymptotic regions found in a universe or multiverse and are solutions to Einstein's field equations. These objects have many interesting features as far as physics is concerned. Morris and Thorne introduced traversable wormholes, which increases the possibility of space-time travel. In this work, the wave equation of the Morris-Thorne wormhole has been derived by the technique of differential forms. The solution of the wave equation for a particular choice of red-shift function and shape function is obtained. The potential has also been computed in order to analyze electromagnetic fields. The behavior of electromagnetic multipoles is expressed and investigated in their behavior at the wormhole's throat.}

\keywords{Debye's Potential, Wormhole, Differential Forms, Multipoles.}



\maketitle

\section{Introduction}\label{sec1}

To study electromagnetism, it is necessary to solve Maxwell's field equations for the electromagnetic fields $\overrightarrow{E}$ and $\overrightarrow{B}$. Maxwell's equations form a system of eight nonlinear partial differential equations with six unknowns. In a flat space, it can be solved by standard methods. In the presence of gravity, the spacetime is curved hence in such a case one needs to solve Maxwell’s equations in curved space, where it is very complicated to solve these equations due to their coupled nature. Also, since the curvature of space-time is attributed to gravitation in the general theory of relativity, the situation becomes more complicated when the interaction of electromagnetism with gravitation is considered. Electromagnetic fields can be given in the form of Maxwell's electromagnetic field tensor $F_{ij}$ and an electromagnetic field tensor can be formed using potentials.
In this article, we have followed terminologies used in \cite{gri, stkmh}.
 \\
\indent The potential formalism is well known in electrodynamics in flat spaces. Due to the coupled structure of Maxwell's equations, there is no standard way to solve them for a curved spacetime, and the approach has to be heuristic. Following are the two most popular techniques, which are introduced by Cohen and Kegeles \cite{candke}:
\begin{enumerate}
       \item Debye's Potential Formalism,
        \item  Newman-Penrose (NP) Formalism.
        \end{enumerate}
 \indent While studying electric dipole fields Hertz introduced the potential of the Maxwell field. Laporte and Uhlenbeck \cite{lap} considered the true covariant bivector nature of the potential.
   Hertz bivector has been reduced to two purely radial bivectors (Debye's potential) by Nisbet \cite{nis}.
 Cohen and Kegeles \cite{candke} have found Debye's potential in the simpler situation of Schwarzschild, Kerr and Friedmann spacetimes and analyzed electromagnetic multipoles.
 In Newmann-Penrose notations, Castillo \cite{cas} found Debye's potential for self-dual fields. Cohen et al. \cite{cvd} used Debye's formalism to study electromagnetism in G\"{o}del spacetime.
 Using Newmann-Penrose (NP) formalism, Hasmani and Patel \cite{hb} found Debye's potential for Vaidya space-time and studied electromagnetic field.
 Hasmani et al. \cite{ha} obtained Debye's potential for pp-waves spacetime and analyzed the interaction of electromagnetism and gravity for it with the help of Newmann-Penrose (NP) formalism. By Geroch-Held-Penrose (GHP) formalism David \cite{dav} discussed the possible relation between the Debye potentials and Lunin potential.
 Teukolsky-Starobinsky identities in terms of the Debye's potentials are obtained by David \cite{dav1}. Kang and Kim investigated the gravitational perturbation of the Morris-Thorne wormhole using the Newmann Penrose formalism \cite{kang}. \\
 \indent In this paper to study electromagnetism in curved spacetime, we have obtained Debye's potential for a wormhole metric and found multipoles for the electric field and magnetic field. We have studied these multipoles at the throat of the wormhole.

\section{Debye's Potential and Differential Forms}\label{sec2}

In this section, we will describe the formalism set up by Cohen and Kegeles \cite{candke} which will be in part used. Also, some necessary changes will be discussed during the usage of their occurrence.\\
 \indent The equations of the two scalar potentials in relation to electric and magnetic polarizations $\vec{P}$ and $\vec{M}$ are given by
\begin{align}\label{1}
\nabla^2\vec{P_E}-\frac{\partial^2 \vec{P_E}}{\partial t^2}=4\pi \vec{P}, \hspace{0.5cm}
\nabla^2\vec{P_M}-\frac{\partial^2 \vec{P_M}}{\partial t^2}=4\pi \vec{M},
\end{align}
where $\vec{P}_{E}$ and  $\vec{P}_{M}$ are the electric and magnetic bivectors of Hertz potential respectively. A second-order derivative of the bivector potential yields the physical fields \cite{candke}. The relationships between potentials and the quantities represented by electric and magnetic bivectors are as follows:
\begin{align}\label{2}
\phi=-\vec{\nabla} \cdot \vec{P}_{E,}\hspace{0.5cm}
\vec{A}=\frac{\partial\vec{P}_{E}}{\partial t}+\vec{\nabla}\times \vec{P}_{M}.
\end{align}
Also, the connections between the electric and magnetic fields and electric and magnetic bivectors are given by,
\begin{align}\label{3}
\vec{E}&=\vec{\nabla}\vec{\nabla}\cdot \vec{P_{E}}-\frac{\partial^2\vec{P_{E}}}{\partial t^2}- \vec{\nabla} \times \frac{\partial\vec{P_{M}}}{\partial t}\nonumber\\
&=-\vec{\nabla} \times \frac{\partial\vec{P_{M}}}{\partial t}+\vec{\nabla}\times(\vec{\nabla}\times\vec{P_{E}}),\nonumber\\
\overrightarrow{B}&=\vec{\nabla} \times \frac{\partial\vec{P_{E}}}{\partial t}+\vec{\nabla}\times(\vec{\nabla}\times\vec{P_{M}})\nonumber\\
&=\vec{\nabla}\vec{\nabla}\vec{P_{M}}-\frac{\partial^2\vec{P_{M}}}{\partial t^2}- \vec{\nabla} \times \frac{\partial\vec{P_{E}}}{\partial t}.
\end{align}
The notations are standard and here $c=1$ is considered. Maxwell's equations in the source-free medium are
\begin{align}\label{4}
\vec{\nabla}\times \vec{E}+\frac{\partial \vec{B}}{\partial t}&=0,\hspace{0.5cm}
\vec{\nabla} \cdot \vec{B}=0,\nonumber\\
\vec{\nabla}\times \vec{B}-\frac{\partial \vec{E}}{\partial t}&=0,\hspace{0.5cm}
\vec{\nabla} \cdot \vec{E}=0
\end{align}
and then equations (\ref{3}) reduce to wave equations for $\vec{P}_{E}$ and $\vec{P}_{M}$
\begin{align}\label{5}
\Box\vec{P}_{E}=0, \hspace{0.5cm}
\Box\vec{P}_{M}=0.
\end{align}
The uniqueness of the Hertz potential is achieved by a new gauge condition. This was termed as gauge transformation of the third kind by Nisbet \cite{nis}. In the wave equations (\ref{5}) the gauge appears as the source.
 This is the source-preserving free property of Maxwell's equation, which become
\begin{align}\label{6}
\vec{Q}_{E}&=\vec{\nabla}\times \vec{G},\hspace{0.5cm} \vec{Q}_{M}=-\frac{\partial \vec{G}}{\partial t}-\vec{\nabla}g, \nonumber\\
\vec{R}_{E}&=-\frac{\partial \vec{W}}{\partial t}-\vec{\nabla}w, \hspace{0.5cm} \vec{R}_{M}=-\vec{\nabla} \times \vec{W},
\end{align}
where $(G,g)$ and $(W,w)$ are arbitrary vectors. In this formalism, the potentials and the wave equations are modified as \cite{candke},
\begin{align}\label{7}
\vec{E}&=\vec{R_E}+\vec{\nabla}\vec{\nabla}\cdot \vec{P_{E}}-\frac{\partial^2\vec{P_{E}}}{\partial t^2}- \vec{\nabla} \times \frac{\partial\vec{P_{M}}}{\partial t}\nonumber\\
&=-\vec{Q}_{E}-\vec{\nabla}\times\frac{\partial \vec{P}_{M}}{\partial t}+\vec{\nabla}\times (\vec{\nabla}\times\vec{P}_{E}),\nonumber\\
\vec{B}&=-\vec{R_M}+\vec{\nabla} \times \frac{\partial\vec{P_{E}}}{\partial t}+\vec{\nabla}\times(\vec{\nabla}\times\vec{P_{M}})\\
&=\vec{Q}_{M}+\vec{\nabla}\vec{\nabla}\cdot \vec{P}_{M}-\frac{\partial^2 \vec{P}_{M}}{\partial t^2}+\vec{\nabla}\times \frac{\partial \vec{P}_{E}}{\partial t}.\nonumber
\end{align}
and
\begin{align}\label{8}
\Box\vec{P}_{E}&=\vec{Q}_{E}+\vec{R}_{E}, \nonumber \\ \Box\vec{P}_{M}&=\vec{Q}_{M}+\vec{R}_{M}.
\end{align}

Cohen and Kegeles \cite{candke} presented all the above equations in differential forms. Here, we consider $d$ as the exterior derivative and $\delta= *d*$, where $*$ denotes Hodge dual \cite{hodge}. Also, we use harmonic operator, that is, $\Delta=d\delta+\delta d$.\\
 \indent In terms of the Maxwell 2-form, electromagnetic field tensor is given by
\begin{align}\label{9}
F=\frac{1}{2}F_{ab}\omega^a \wedge \omega^b,
\end{align}
where $F_{ab}$ denote components of Maxwell 2-form and $\omega^a$ are basis forms. Here, $a$ and $b$ in what follows the Latin indices have range 0, 1, 2, 3. This convention of range of Latin indices will be followed through out the paper. The Maxwell equations (\ref{4}) become
\begin{align}\label{10}
dF&=0,\\
\delta F&=0.
\end{align}
 The equations (\ref{2}) relating Hertz bivector (2-form) $P$ to the four-vector (1-form) potential $A$ become
\begin{align}\label{11}
A=\delta P.
\end{align}
The analogue of equations (\ref{3}) giving $F$ in the terms of $P$ is
\begin{align}\label{12}
F=d \delta P=-\delta dP.
\end{align}
The equality of the last two expressions in (\ref{12}) requiring
\begin{align}\label{13}
\Delta P=0,
\end{align}
which is an analogue of equations (\ref{5}). The exterior derivative operator $d$ satisfies $d^2=0$. Thus,
\begin{align*}
dF=d(d\delta P)=0,\\
\delta F= \delta(-\delta dP)=0.
\end{align*}
 The 2-form gauge terms (\ref{6}) are
\begin{align}\label{14}
Q=dG, \hspace{0.3cm}
R=*dW,
\end{align}
where $G$ and $W$ are arbitrary 1-forms. The wave equation (\ref{8}) with gauge terms is therefore
\begin{align}\label{15}
\Delta P=dG+*dW,
\end{align}
so that the gauge-transformed field (\ref{7}) becomes
\begin{align}\label{16}
F=d \delta P-dG=*dW-\delta dP.
\end{align}
 Equations (\ref{15}) and (\ref{16}) represent a fully covariant generalization of the Hertz 2-vector (Debye's) potential scheme to all curved spacetimes.

\section{Debye's Potential for Morris-Thorne Wormhole}\label{sec3}
The metric for  Morris-Thorne wormhole \cite{mtw} is given by,
\begin{eqnarray}\label{18}
ds^2 = -e^{2\Phi(r)}dt^{2}+\frac{dr^2}{\left(1-\frac{b(r)}{r}\right)}+r^2d\theta^2+r^2\sin^2\theta d\phi^2,
\end{eqnarray}
In the context of wormhole solutions, $\Phi(r)$ represents the red-shift function and $b(r)$ represents the shape function, it is crucial to ensure the absence of an event horizon within the traversable wormhole. Additionally, minimizing the impact of tidal gravity forces on the traveler is essential. To establish the conditions for the existence of these wormhole solutions, the shape function must adhere to the following criteria: at a specific point $r_0$, the shape function satisfies $b(r_0) = r_0$.
The flare-out condition, denoted as $\frac{b(r)-b'(r)r}{b^2(r)}>0$, must hold for all $r > r_0$.
The derivative of the shape function, $b'(r)$, remains below 1.
For values of $r$ greater than $r_0$, the ratio $\frac{b(r)}{r}$ remains less than 1.
As the distance $r$ approaches infinity, the ratio $\frac{b(r)}{r}$ tends to zero. Also, the red-shift function should be finite everywhere.\\
For orthonormal Cartan frame  \cite{isrl} we choose following 1-forms,
\begin{align}\label{19}
\omega^0&=e^{\Phi(r)}dt, \hspace{2.8cm}   \omega^2=rd\theta, \nonumber \\
\omega^1&=\left(1-\frac{b(r)}{r}\right)^{-1/2} dr, \hspace{1cm} \omega^3=r\sin\theta d\phi.
\end{align}
 Hertz 2-form \cite{guv, stjm} can be chosen as
\begin{equation}\label{20}
  P=P_E \omega^0 \wedge \omega^1 + P_M \omega^2 \wedge \omega^3,
\end{equation}
where $P_E$ and $P_M$ are electric and magnetic  Debye's potentials respectively.
Since Hodge dual $*$ and Harmonic operator $\Delta$ commute and by characteristics of Hodge dual, we choose $P_E=\Psi$ and $P_M=0$ (or $P_M=\Psi$ and $P_E=0$ can also be chosen) for further ease of computation.
 By applying the Harmonic operator on $P$. That is,
\begin{align}\label{21}
\Delta P=d\delta P+\delta dP,
\end{align}
we get an equation,
\begin{align}\label{22}
  \Delta P&=[e^{-2\Phi(r)}\Psi_{,tt}-[(\Psi r^2)_{,r}\left(1-\frac{b(r)}{r}\right)^{1/2}r^{-2}e^{\Phi(r)}]_{,r} e^{-\Phi(r)}\left(1-\frac{b(r)}{r}\right)^{1/2}-\Psi_{,\phi\phi}(r^2\sin^2\theta)^{-1}\nonumber\\
  &\hspace{0.5cm}-r^{-2}\Psi_{,\theta\theta}
  -r^{-2}\Psi_{,\theta}\cot\theta]\omega^0 \wedge \omega^1
  -[2\Psi_{,\theta} r^{-2}\left(1-\frac{b(r)}{r}\right)^{1/2}]\omega^0 \wedge \omega^2\nonumber\\
  &\hspace{0.5cm}-[2\Psi_{,\phi}(r^2\sin\theta)^{-1}\left(1-\frac{b(r)}{r}\right)^{1/2}]\omega^0 \wedge \omega^3,
\end{align}
where a suffix followed by a comma($,$) denotes partial differentiation with respect to the respective variable.\\
Since potentials are unique up to gauge conditions, we choose gauge terms as,
 \begin{align}\label{23}
   G=\frac{2P_E}{r}\left(1-\frac{b(r)}{r}\right)^{1/2}\omega^0 \hspace{0.4cm}\text{and} \hspace{0.4cm} W=\frac{2P_M}{r}\left(1-\frac{b(r)}{r}\right)^{1/2}\omega^0.
 \end{align}
 Now using,
 \begin{equation}\label{24}
   \Delta P=dG+*dW,
 \end{equation}
 we get wave equation for each of $P_E$ and $P_M$,
 \begin{align}\label{25}
   e^{-2\Phi(r)}\Psi_{,tt}-[(\Psi)_{,r}\left(1-\frac{b(r)}{r}\right)^{1/2}e^{\Phi(r)}]_{,r} e^{-\Phi(r)}\left(1-\frac{b(r)}{r}\right)^{1/2} \nonumber \\
   -\Psi_{,\phi\phi}(r^2\sin^2\theta)^{-1}
  -r^{-2}\Psi_{,\theta\theta}
  -r^{-2}\Psi_{,\theta}\cot\theta&=0.
 \end{align}
  By solving the above wave equation, Debye's potential can be determined and it is very difficult to solve the above equation.\\
 \indent For the simple Morris-Thorne wormhole, a red-shift function is $\Phi(r)=0$ and shape function $b(r)=\frac{r_0^2}{r}$ \cite{kang} can be chosen. So, the wave equation becomes
 \begin{align}\label{26}
 &\Psi_{,tt}-[(\Psi)_{,r}\left(1-\frac{r_0^2}{r^2}\right)^{1/2}]_{,r} \left(1-\frac{r_0^2}{r^2}\right)^{1/2} \nonumber\\
  &-\Psi_{,\phi\phi}(r^2\sin^2\theta)^{-1}
  -r^{-2}\Psi_{,\theta\theta}
  -r^{-2}\Psi_{,\theta}\cot\theta=0.
 \end{align}
By the technique of separation of variables, the solution of the above wave equation (\ref{26}) is of the form
\begin{align}\label{27}
\Psi=e^{-ikt}R(r)Y^n_m(\theta, \phi),
\end{align}
which is a Debye's potential for the simplest Morris-Thorne wormhole (\ref{18}).\\
Here, $R(r)$ is a solution of
\begin{align}\label{28}
  k^2R+\frac{1}{A}\frac{d}{dr}\left[\frac{1}{A}\frac{dR}{dr}\right]-\frac{m(m+1)}{r^2}R=0,
  \end{align}
  where $A=\left(1-\frac{r_0^2}{r^2}\right)^{1/2}$. In equation (\ref{28}) there are singularities namely, $r=\pm r_0$ and $r=0$. \\
To solve this differential equation the substitution $\left(1-\frac{r_0^2}{r^2}\right)=x^2$ reduces equation (\ref{28}) to,
 \begin{align}\label{x}
 (1-x^2)^3\frac{d^2R}{dx^2}-3x(1-x^2)^2\frac{dR}{dx}+[r_0^2k^2-m(m+1)(1-x^2)]R=0
 \end{align}
 In the equation (\ref{x}), $x=\pm 1$
are singular points where the condition of the throat of a wormhole is violated and hence the approach of power series can be employed, thus  we get a solution of (\ref{x}) is of the form,

\begin{align}\label{x1}
R(x)&=C_0\left[1+\frac{1}{2}(m(m+1)-r_0^2k^2)x^2\right. \nonumber\\
&\hspace{0.5cm}\left.+\left(\frac{1}{24}(m(m+1)-r_0^2k^2)(12+m(m+1)-r_0^2k^2)-\frac{1}{12}m(m+1)\right)x^4+...\right]\nonumber\\
&\hspace{0.5cm}C_1\left[x+\frac{1}{6}(3+m(m+1)-r_0^2k^2)x^3\right. \nonumber\\
&\hspace{0.5cm}\left.+\frac{1}{120}\left(\frac{1}{120}(3+m(m+1)-r_0^2k^2)(27+m(m+1)-r_0^2k^2)-\frac{1}{20}(6+m(m+1))\right)x^5+...\right]
\end{align}
and solution it the terms of radial coordinate $r$ is given by,

 \begin{align}\label{29}
 R(r)&=C_0\left[1+\frac{1}{2}(m(m+1)-r_0^2k^2)\left(1-\frac{r_0^2}{r^2}\right)\right. \nonumber\\
&\hspace{0.5cm}\left.+\left(\frac{1}{24}(m(m+1)-r_0^2k^2)(12+m(m+1)-r_0^2k^2)-\frac{1}{12}m(m+1)\right)\left(1-\frac{r_0^2}{r^2}\right)^2+...\right]\nonumber\\
&\hspace{0.5cm}C_1\left[\left(1-\frac{r_0^2}{r^2}\right)^{1/2}+\frac{1}{6}(3+m(m+1)-r_0^2k^2)\left(1-\frac{r_0^2}{r^2}\right)^{3/2}\right. \nonumber\\
&\hspace{0.5cm}\left.+\frac{1}{120}\left(\frac{1}{120}(3+m(m+1)-r_0^2k^2)(27+m(m+1)-r_0^2k^2)\right. \right. \nonumber\\
&\hspace{0.5cm}\left. \left.-\frac{1}{20}(6+m(m+1))\right)\left(1-\frac{r_0^2}{r^2}\right)^{5/2}+...\right]
 \end{align}
where $C_0$ and $C_1$ arbitrary constants and $Y^n_m(\theta,\phi)$ are spherical harmonics
 \begin{align}\label{30}
 Y^n_m(\theta,\phi)=\sqrt{\frac{2m+1}{4\pi}\frac{(m-n)!}{(m+n)!}}P^n_m(\cos \theta)e^{in\phi},
 \end{align}
where $P^n_m(\cos \theta)$ is an associated Legendre polynomial.

\section{Electromagnetic Multipoles}\label{sec4}
Referring to Cohen and Kegeles \cite{candke}, we can see that they have not given the form of the fucntion $R$ and hence numerical method was used for analysis. In our case since we could find an analytic solution though it is not given in the closed form. Now we will use the solution of (\ref{28}) for a discussion of electromagnetic multipoles applying appropriate truncation.\\
\indent In terms of the basis 2-forms the tetrad form of electromagnetic field tensor is given by equation (\ref{9}) with the physical correspondences,
\begin{align}\label{31}
E_j=F_{j0}, \hspace{0.5cm} B_1=F_{23}, \hspace{0.5cm} B_2=F_{31}, \hspace{0.5cm} B_3=F_{12},
\end{align}
where $j=1, 2, 3.$
 By equation (\ref{16}) we get,
 \begin{align}\label{32}
 F&=[\Psi_{,tt}-A^{-1}(\Psi_{,r}A^{-1})_{,r}]\omega^0 \wedge \omega^1-\Psi_{,r\theta}(rA)^{-1} \omega^0 \wedge \omega^2- \Psi_{,r\phi}(rA\sin\theta)^{-1} \omega^0 \wedge \omega^3 \nonumber \\
 &\hspace{0.5cm}-\Psi_{,t\theta}r^{-1} \omega^1 \wedge \omega^2-\Psi_{,t\phi}(r\sin\theta)^{-1} \omega^1 \wedge \omega^3.
 \end{align}
 Using equations (\ref{16}), (\ref{31}), and (\ref{32}),
 \begin{align}
 E_1&=\frac{1}{A} \frac{\partial}{\partial r}\left(\frac{1}{A} \frac{\partial \Psi}{\partial r}\right)-\frac{\partial^2 \Psi}{\partial t^2}.
 \end{align}
 Using equation (\ref{28}),
 \begin{align}
 E_1&= \left[\frac{m(m+1)}{r^2}\right]e^{-ikt}R(r)Y^n_m(\theta,\phi).
 \end{align}
In a similar manner we get electric multipoles as follows:
\begin{align}\label{x2}
E_1&= \left[\frac{m(m+1)}{r^2}\right]e^{-ikt}R(r)Y^n_m(\theta,\phi), \hspace{0.4cm} B_1=0,\nonumber\\
E_2&=\frac{e^{-ikt}}{Ar}\frac{\partial}{\partial r}[R(r)]\frac{\partial}{\partial \theta}[Y^n_m(\theta,\phi)], \hspace{0.4cm} B_2=\frac{kne^{-ikt}}{r\sin\theta} R(r)Y^n_m(\theta,\phi),\nonumber\\
E_3&=\frac{ine^{-ikt}}{Ar\sin\theta}\frac{\partial}{\partial r}[R(r)]Y^n_m(\theta,\phi), \hspace{0.4cm} B_3=\frac{ike^{-ikt}}{r}R(r)\frac{\partial}{\partial \theta}[Y^n_m(\theta,\phi)].
\end{align}
 These are the electric multipoles except for $m=0$ in $E_1$, and by performing duality operation we get magnetic multipoles which are given by,
 \begin{align}
B_1&= \left[\frac{m(m+1)}{r^2}\right]e^{-ikt}R(r)Y^n_m(\theta,\phi), \hspace{0.4cm} E_1=0,\nonumber\\
B_2&=\frac{e^{-ikt}}{Ar}\frac{\partial}{\partial r}[R(r)]\frac{\partial}{\partial \theta}[Y^n_m(\theta,\phi)], \hspace{0.4cm} E_2=-\frac{kne^{-ikt}}{r\sin\theta} R(r)Y^n_m(\theta,\phi),\nonumber\\
B_3&=\frac{ine^{-ikt}}{Ar\sin\theta}\frac{\partial}{\partial r}[R(r)]Y^n_m(\theta,\phi), \hspace{0.4cm} E_3=-\frac{ike^{-ikt}}{r}R(r)\frac{\partial}{\partial \theta}[Y^n_m(\theta,\phi)].\nonumber
\end{align}
We observe that the electric and magnetic multipoles are nothing but the corresponding components of electromagnetic field tensor and dual of electromagnetic field tensor and they are derived using 1-forms. this demonstrates an application of simplifying the computation.\\
\indent In the static situation, $k=0$, and at the wormhole's throat, $R(r)=C_0$, hence the only non-zero component of electric multipoles is $E_1$, which is represented as,
\begin{align}\label{x3}
E_1&= C_0\left[\frac{m(m+1)}{r_0^2}\right]Y^n_m(\theta,\phi)=F_{10}.
\end{align}
The non-zero component of magnetic multipoles is $B_1$,
\begin{align}\label{x4}
B_1&= C_0\left[\frac{m(m+1)}{r_0^2}\right]Y^n_m(\theta,\phi)
\end{align}
and out of throat expressions of non-zero components of electric multipoles are $E_1, E_2, E_3$ same as given in equation (\ref{x2}), Similar in the case for magnetic multipoles.

\section{Conclusion}\label{sec5}

The wave equation corresponding to the Morris-Thorne wormhole is established through a methodology of Cohen and Kegeles. This equation allows us to deduce Debye's potential for a specific scenario where the red-shift function is taken as zero and the shape function follows a pattern of $\frac{r_0^2}{r}$. Consequently, electromagnetic multipoles for the Morris-Thorne wormhole are computed. While examining electric multipoles, it's evident that the $B_1$ component becomes null, signifying that electric multipole fields take the form of spherical transverse magnetic waves or spherical electric waves. Similarly, for magnetic multipoles, the $E_1$ component is found to be zero, resulting in spherical transverse electric waves or spherical magnetic waves.\\
\indent Furthermore, electromagnetic multipoles are calculated for the static condition at the wormhole's throat. In this scenario, only one non-zero component, $E_1$, emerges, which relies on both $r_0$ and a spherical harmonic function. As the throat's radius increases significantly, all the multipoles gradually vanish. Beyond the throat, only components of the electric field are observed. Thus, the gravitational field described by the Morris-Thorne wormhole exhibits purely electric characteristics. Recently, Soni et. al. \cite{soni} have shown that the Morris-Thorne wormhole is of Petrov type D. This analysis demonstrates that in both instances, whether at the throat or beyond, the only active components of the electromagnetic field are $F_{j0}$ for $j=1, 2, 3$, corresponding to the electric field. As a result, magnetic field interaction is absent in the Morris-Thorne wormhole which matches with the result ``Any Petrov type D spacetimes is either purely electric or purely magnetic". The influence of the higher powers of $\frac{1}{r}$ in the expression of electromagnetic multipoles appears to overshadow the impact of gravitational force at the throat, thereby reinforcing the throat's characteristic features in a wormhole.

\bmhead{Acknowledgments}

SVS is thankful to the CSIR, India for providing financial support under CSIR Senior Research Fellowship (09/157(0059)/2021-EMR-I). RP is thankful to Sardar Patel University for providing financial support under SEED Money Grant. The authors are also thankful to Dr. Meera H. Chudasama for her help in solving differential equations.

\section*{Declarations}

\bmhead {Competing interests} The authors declare no competing interests.

\bibliography{sn-bibliography}

\end{document}